\documentclass[twocolumn,english,aps,prb,showpacs,superscriptaddress,floats,amsmath,amssymb,floatfix]{revtex4}
\usepackage[latin9]{inputenc}
\usepackage{color}
\usepackage{amsmath}
\usepackage{graphicx}
\usepackage{graphics}
\usepackage[pdf]{pstricks} 
\usepackage{pdftexcmds}
\usepackage{ifpdf}
\usepackage{amssymb}
\usepackage{esint}

\makeatletter
\@ifundefined{textcolor}{}
{%
 \definecolor{BLACK}{gray}{0}
 \definecolor{WHITE}{gray}{1}
 \definecolor{RED}{rgb}{1,0,0}
 \definecolor{GREEN}{rgb}{0,1,0}
 \definecolor{BLUE}{rgb}{0,0,1}
 \definecolor{CYAN}{cmyk}{1,0,0,0}
 \definecolor{MAGENTA}{cmyk}{0,1,0,0}
 \definecolor{YELLOW}{cmyk}{0,0,1,0}
 }

\usepackage{array}
\usepackage{babel}\@ifundefined{definecolor}
 {\usepackage{color}}{}

\allowdisplaybreaks

\makeatother

\usepackage{babel}

\begin{document}

\title{Quantum phases of a frustrated four-leg spin tube}

\author{M.~Arlego}

\email{arlego@fisica.unlp.edu.ar}

\affiliation{Departamento de F\'{i}sica, Universidad Nacional de La Plata, C.C.
67, 1900 La Plata, Argentina}

\author{W.~Brenig}

\affiliation{Institut für Theoretische Physik, Technische Universität Braunschweig,
38106 Braunschweig, Germany }

\author{Y.~Rahnavard}

\affiliation{Institut für Theoretische Physik, Technische Universität Braunschweig,
38106 Braunschweig, Germany }

\author{B.~Willenberg}

\affiliation{Institut für Theoretische Physik, Technische Universität Braunschweig,
38106 Braunschweig, Germany }

\author{H.D.~Rosales }

\affiliation{Departamento de F\'{i}sica, Universidad Nacional de La Plata, C.C.
67, 1900 La Plata, Argentina}

\author{G.~Rossini}

\affiliation{Departamento de F\'{i}sica, Universidad Nacional de La Plata, C.C.
67, 1900 La Plata, Argentina}
\begin{abstract}
We study the ground state phase diagram of a frustrated spin-1/2 four-leg
tube. Using a variety of complementary techniques, namely density matrix
renormalization group, exact diagonalization, Schwinger boson mean field theory,
quantum Monte-Carlo and series expansion, we explore the parameter space of
this model in the regime of all-antiferromagnetic exchange. In contrast to
unfrustrated four-leg tubes we uncover a rich phase diagram. Apart from the
Luttinger liquid fixed point in the limit of decoupled legs, this comprises
several gapped ground states, namely a plaquette, an incommensurate, and an
antiferromagnetic quasi spin-2 chain phase. The transitions between these phases
are analyzed in terms of total energy and static structure factor calculations
and are found to be of (weak) first order. Despite the absence of long
range order in the quantum case, remarkable similarities to the classical phase
diagram are uncovered, with the exception of the icommensurate regime, which is
strongly renormalized by quantum fluctuations. In the limit of large leg
exchange the tube exhibits a deconfinement cross-over from gapped magnon like
excitations to spinons.
\end{abstract}


\maketitle

\section{Introduction}

Quasi one-dimensional spin systems, comprising chain, ladder and more
involved magnetic structures are an active field of research thriving
on a constant feedback between material synthesis, experimental investigations
and theoretical predictions \cite{Dagotto1996a,Lemmens2003,Batchelor2007}.

Magnetic frustration is a key issue in this field, which has experienced
an upsurge of interest, starting with the discovery of $J_{1}$-$J_{2}$
chain materials, like CuGeO$_{3}$ \cite{Hase93}, followed by the
investigation of spin tube compounds with an odd number $N$ of sites
per unit cell, such as {[}(CuCl$_{2}$tachH)$_{3}$Cl{]}Cl$_{2}$
\cite{Schnack2004} and CsCrF$_{4}$ \cite{Manaka2009} with $N=3$,
and Na$_{2}$V$_{3}$O$_{7}$ \cite{Millet1999} with $N=9$. Spin
tubes with an odd number of legs and only nearest neighbor antiferromagnetic
(AFM) exchange are \emph{geometrically} frustrated. Because of the
Lieb-Schultz-Mattis theorem the ground state of such systems is either
gapless and non-degenerate, or gapped with a broken translational
invariance. Indeed, for spin-1/2 tubes with $N=3$ a spin gap was
found in case of identical couplings on the triangular rungs, with
a transition into a gapless and translationally invariant phase at
already weakly non-equivalent couplings.\cite{Fouet2006a,Nishimoto2008a,Ivanov2010,Sakai2010,Pujol2012}
$N=3$ spin-1/2 tubes with isosceles triangle basis also show a 1/3
magnetization-plateau.\cite{Sato2007}

Recently, Cu$_{2}$Cl$_{4}$$\cdot$D$_{8}$C$_{4}$SO$_{2}$ has
been established as a new spin-1/2 tube with an even number of legs
\cite{Garlea2008a}, namely $N=4$. Tubes with $N=4$ and only nearest
neighbor AFM exchange are \emph{not} frustrated. However, substantial
next-nearest neighbor AFM exchange, diagonally coupling adjacent legs,
has been claimed for Cu$_{2}$Cl$_{4}$ $\cdot$D$_{8}$C$_{4}$SO$_{2}$,
rendering also this ladder system frustrated. Inelastic neutron scattering
\cite{Zheludev2008a,Garlea2009} has revealed a strongly one-dimensional
(1D) elementary excitation, which is gapped and slightly incommensurate.
The former is consistent with Haldane's conjecture \cite{Haldane1983}
for 1D spin systems with an even number of spin-1/2 moments per unit
cell. The latter is consistent with a frustrated exchange. Magnetic
fields have been shown to stabilize the incommensurate spin correlations.\cite{Zheludev2008a,Garlea2009}

\begin{figure}[tb]
\begin{centering}
\includegraphics[width=0.7\columnwidth,angle=-90]{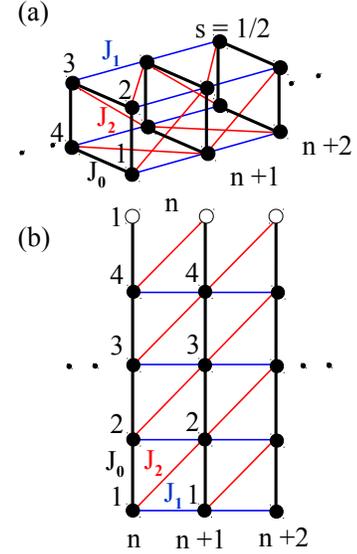}
\vspace{2cm}
\par\end{centering}

\caption{\label{fig1} (color online) (a) Frustrated four-spin tube. Solid circles represent
spin-$1/2$ moments. Plaquettes (bold black lines) are coupled by nearest
($J_{1}$) and next nearest ($J_{2}$) antiferromagnetic exchange,
blue and red lines, respectively. On-plaquette coupling is $J_{0}$.
(b) Frustrated four-spin tube unwrapped, displaying structure of an
anisotropic triangular lattice on a torus.}
\end{figure}

Motivated by this, a \emph{geometrically} frustrated and simplified
four-spin tube (FFST) model has been introduced in Ref.~\onlinecite{Arlego2011}
\begin{equation}
H=\sum_{lm}J_{lm}\mathbf{S}_{l}\cdot\mathbf{S}_{m}\,,\label{eq:1}\end{equation}
with a lattice structure and exchange couplings $J_{lm}$ as shown
in Fig.~\ref{fig1}. Spin-1/2 moments are located on the solid circles
and all couplings $J_{0,1,2}$ are antiferromagnetic (AFM). The FFST
lattice is identical to an anisotropic triangular lattice on a torus
with four site circumference.

For $J_{1,2}\ll J_{0}$, the quantum properties of the FFST can be
understood in terms of weakly coupled four-spin plaquettes, which allows
for series expansion in terms of $J_{1,2}$. In Ref.~\onlinecite{Arlego2011}
such a series expansion has been carried out in detail regarding the
one- an two-particle excitations in this restricted parameter regime.
However, an understanding of the \emph{quantum} phases of the FFST
on a larger scale is still missing.

Therefore, in this paper we will present combined results from a large
variety of complementary methods, namely, density-matrix renormalization
group (DMRG), exact diagonalization (ED), series expansion (SE), Schwinger
bosons mean field theory (SBMFT), and quantum Monte-Carlo (QMC) in
order to explore the parameter space of the FFST.
We set $J_0=1$, except where explicitly indicated,
and denote by $L$ the tube length.

The structure of the paper is as follows. In section \ref{sec:Classical-phase-diagram}
we briefly summarize the phase diagram of the classical FFST. In section
\ref{sec:Quantum-phases-and} we consider the quantum case focusing
the discussion onto the strong and intermediate on-plaquette exchange
in subsection \ref{SIC} and on the limit of very large leg exchange
in subsection \ref{LWC}. Section \ref{sec:Discussion-and-Conclusions}
summarizes our picture of the FFST. For completeness we briefly summarize
some of the methods used and refer to important references for them
in appendix \ref{sec:Techniques}.


\section{Classical phase diagram\label{sec:Classical-phase-diagram}}

\begin{figure}[tb]
\begin{centering}
\includegraphics[width=0.8\columnwidth]{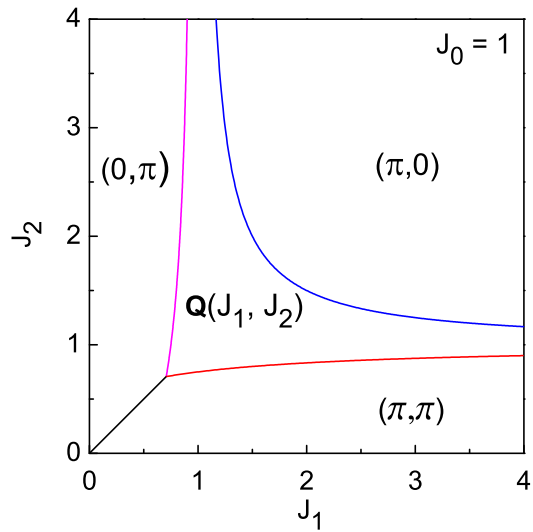}
\par\end{centering}

\caption{\label{fig2} (color online) Four classical phases of the FFST with commensurate
pitches $\mathbf{Q}=(\pi,\pi)$, $(\pi,0)$, $(0,\pi)$ and incommensurate
regime $\mathbf{Q}(J_{1},J_{2})$ as in text.}
\end{figure}

While we are studying a quantum model in one dimension which does
not allow for breaking of a continuous symmetry at zero temperature,
it is nevertheless instructive to compare the quantum case considered
in the following sections with a \emph{classical} magnetic phase diagram
of the FFST. From Ref.~\onlinecite{Arlego2011} it is known that there are
four ordered regimes in which $\mathbf{S}(\mathbf{r}_{l})=S(\cos(\mathbf{Q}\cdot\mathbf{r}_{l}),\sin(\mathbf{Q}\cdot\mathbf{r}_{l}),0)$,
where $\mathbf{r}_{l}$ is a lattice site:
\begin{enumerate}
\item \setlength{\itemsep}{-1mm}$J_{2}\leqslant(1+2J_{1})/(2(J_{1}+1))$ and $J_{2}\leqslant J_{1}$:
commensurate $\mathbf{Q}=(\pi,\pi)$ AFM
\item $J_{2}\geqslant(1-2J_{1})/(2(J_{1}-1))$, with $J_1<1$, and $J_{2}\geqslant J_{1}$:
commensurate $\mathbf{Q}=(0,\pi)$ AFM
\item $J_{2}\geqslant(2J_{1}-1)/(2(J_{1}-1))$, with $J_{1}>1$:
commensurate $\mathbf{Q}=(\pi,0)$ AFM
\item \label{enu:Two-degenerate-incommensurate}Two degenerate incommensurate
spirals with $\mathbf{Q}(J_{1},J_{2})=\pm(2\arctan(\alpha),\pi/2)$,
and $\alpha=(J_{1}+\sqrt{J_{1}^{2}+J_{2}^{2}})/J_{2}$ in the remaining
region.
\end{enumerate}
These are shown in Fig.~\ref{fig2}. All classical transitions are
of first order.


\section{Quantum phases and correlation functions\label{sec:Quantum-phases-and}}

In the following we gather information from various complementary
methods to develop a quantum version of the phase diagram of the FFST.
The dicussion is split into two subsections. The first focuses on
the strong and intermediate on-plaquette exchange and comprises an
analysis of the ground state energy using DMRG, SE, SBMFT, and ED,
followed by an evaluation of the phase diagram from SBMFT, and finally
a DMRG study of correlation funtions and structure factors. In the
second subsection we analyze the spin excitations in the limit $J_{0}\rightarrow0$
using QMC.

To begin, we note that in the quantum case and at the points $J_{1(2)}\rightarrow\infty$,
$J_{2(1)}=0$ the FFST is in a Luttinger liquid (LLQ) state. Staying
on either of the two axes $(J_{1(2)}\neq\infty,J_{2(1)}=0)$ , the
system is unfrustrated, the inter-leg coupling is relevant, and the
FFST opens a spin gap. This gapped phase is adiabatically connected
to that of unfrustrated weakly coupled plaquettes $(J_{1(2)}\ll1,J_{2(1)}=0)$ which
has been studied extensively in Refs.~\onlinecite{Cabra1998,Kim1999}. The
frustrated weakly coupled plaquette regime shows no transition between a $(\pi,\pi)$
and $(0,\pi)$ phase, rendering the diagonal line in the lower left
corner of Fig.~\ref{fig2} a classical-only effect.


\subsection{Strong and intermediate on-plaquette coupling\label{SIC}}


\subsubsection{Ground state energy\label{SSegs}}

\begin{figure}[tb]
\begin{centering}
\includegraphics[width=0.8\columnwidth]{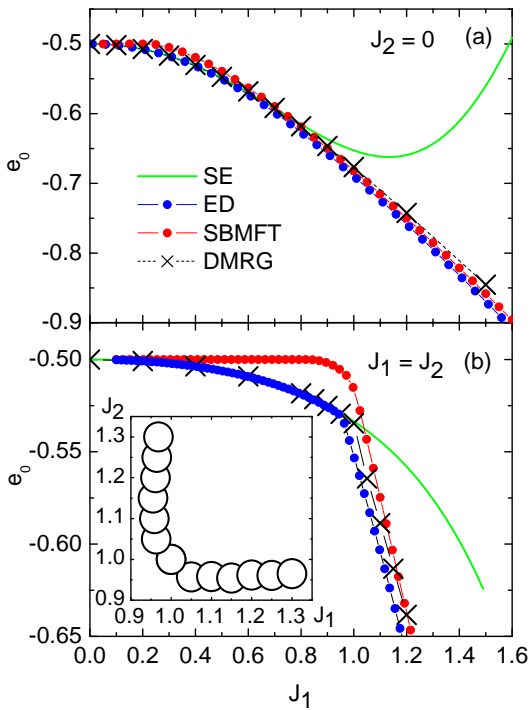}
\par\end{centering}
\caption{\label{fig:egs} (color online) Ground state energy per site $e_0$
for $J_{0}=1$.
Panel (a) $e_0$ vs. $J_{1}$ at $J_{2}=0$.
Panel (b) $e_0$ vs. $J_{2}=J_{1}$.
Green solid: plaquette SE at $O(7)$.
Dotted blue (red): ED with L=6 and PBC (SBMFT with L=400 and PBC).
Large crosses: DMRG with L=20, m=300, and OBC.
Inset in (b): first order transition line
in the $J_{1,2}$-plane.}
\end{figure}

A natural question arising is, how far the weakly coupled plaquette
phase extends away from the $J_{1(2)}$ axes lines and if its break
down is of first or second order. We check this in two ways, considering
the ground state energy $e_{0}$ versus $J_{1,2}$ and the static
structure factor. The results for $e_{0}$ are summarized in
Fig.~\ref{fig:egs}. It depicts the results from different techniques (appendix
\ref{sec:Techniques}), along two paths in parameter space. Panel
(a) is along the $J_{1}$-axis, while panel (b) diagonal path $J_{1}=J_{2}$
of maximum frustration.

Along the $J_{1}$-axis, panel (a) the energy is a smooth function.
All methods are in satisfactory agreement up to $J_{1}\approx0.7$.
At this point the bare SE shown, which has been obtained up to $O(7)$
(appendix \ref{sub:Series-expansion}), loses convergence, while the
other techniques continue to agree throughout the range shown. We
note that finite size effects on the DMRG and ED are expected to be
small since the system is gapped.

Along the line of maximum frustration, Fig.~\ref{fig:egs} (b), the
energy as obtained from DMRG and ED shows an obvious discontinuity
in its first derivative at $J_{1}\approx1$. This signals a first
order quantum phase transition. Remarkably this point is rather close
to the classical tricritical point, separating ($\pi,\pi$), ($0,\pi$)
and spiral classical phases of Fig.~\ref{fig2}. By construction SE
based on a single unperturbed starting state is unable to detect this
transition, which is consistent with Fig.~\ref{fig:egs} (b), where
the SE agrees perfectly with DMRG and ED exactly up to the kink in
$e_{0}$. Finally SBMFT is very close to DMRG and ED in this panel
beyond the transition, however it underestimates the energy severely
at smaller $J_{1}=J_{2}$. We will return to this in subsection \ref{sub:SBMFTphadig}.

Using DMRG ground state energies, we follow the first order transition
in the $J_{1,2}$-plane. This is shown in the inset of
Fig.~\ref{fig:egs} (b).
Apart from
a very small curvature in the immediate vicinity of the transition
point on the diagonal $J_{1}=J_{2}$, the plaquette phase border is
composed of almost straight lines: $J_{2(1)}^{c}(J_{1(2)})\approx1$
for $1\lesssim J_{1(2)}\lesssim1.5$. For values of $J_{1(2)}\gtrsim1.5$,
the error on the detection of the kink from our numerical data is
too large to make definite conclusions. While this is identical to
previous findings in Ref.~\onlinecite{Arlego2011}, our evaluation of the
static structure factor as in subsection \ref{CfSf} shows that the
first order transition is very likely to extend at least up to $J_{1(2)}\approx5$.

In summary, ground state energy calculations seem consistent with
a plaquette phase extending throughout two strips of width of order
unity parallel to each of the $J_{1(2)}$-axis, at least up to intermediate
$J_{1(2)}$. Finally, there are \emph{no} signatures of additional
first order transitions, separating a putative incommensurate and
$(\pi,0)$-phase. In view of this 'missing' second incommensurate-to-commensurate
transition, we will consider also real space correlation functions
and static structure factors in section \ref{CfSf}.


\subsubsection{SBMFT phase diagram\label{sub:SBMFTphadig}}

\begin{figure}
\begin{centering}
\includegraphics[width=0.8\columnwidth]{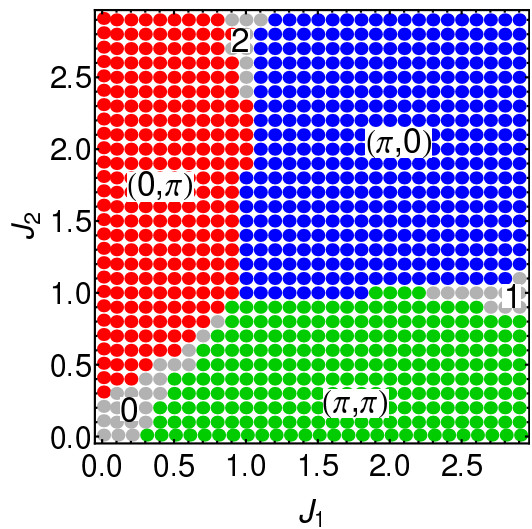}
\par\end{centering}

\caption{\label{fig4} (color online) SBMFT phase diagram. Pitch vectors label short range
spin correlations. Grey regions correspond to unphysical 'decoupled
chain' phases}
\end{figure}

Next we turn to the phase diagram as obtained from SBMFT. We use an
$SU(2)$ invariant decoupling scheme described in appendix \ref{sub:Schwinger-bosons}
focusing on solutions with homogeneous mean fields. Apart from one
Lagrange multiplier to fix the local spin, this leads to \emph{six}
bond parameters $B_{n=0,1,2}$ and $A_{n=0,1,2}$, one $B_n$ and one $A_n$
for each of the three non-equivalent exchange links in Fig.~\ref{fig1}.
$B_{n}$ refer to triplet, and $A_{n}$ to singlet spin correlations.
Solving the self consistency Eqs.~(\ref{w3},\ref{w4}) either
in the continuum limit, or, equivalently minimizing the energy of
Hamiltonian (\ref{w2}) on sufficiently large finite FFSTs, we find
the quantum phase diagram shown in Fig.~\ref{fig4} for $0\leqslant J_{1,2}\leqslant3$.

\begin{figure}
\begin{centering}
\includegraphics[width=0.5\columnwidth]{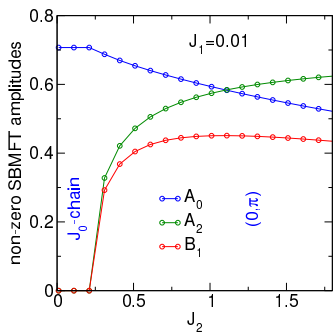}
\par\end{centering}

\caption{\label{fig5} (color online) Non-zero bond mean field parameters within the phases
of Fig.~\ref{fig4} versus $J_{2}$ for $J_{1}=0.01$.}
\end{figure}

First, we emphasize, that the SBMFT solutions in all of the parameter
space investigated remains gapped. I.e., there is no condensation of
Schwinger bosons, and correspondingly no long-range magnetic order
(LRO). This is to be expected in 1D. The 'pitch' vector labels in
Fig.~\ref{fig4} refer to short range correlations as depicted in
Fig.~\ref{fig5}, which shows a vertical cut through the phase diagram
of Fig.~\ref{fig4} close to $J_{1}=0$. In the 'red' phase the AFM
bond mean fields along the plaquette rungs and the diagonal $J_{2}$-links
are finite, while there are ferromagnetic correlations along the $J_{1}$-links.
In this sense this is a $(0,\pi)$-phase, similar to Fig.~\ref{fig2}.
The same notion applies to the $(\pi,0)$- and $(\pi,\pi)$-phase.
All transitions between red, green, and blue phases in Fig.~\ref{fig4}
are of first order.

Fig. \ref{fig5} clearly shows, that upon lowering $J_{2}$ the FFST
continuously evolves into a weakly coupled plaquette regime in the
red phase. I.e., for $J_{2}\lesssim1$, the singlet amplitudes $A_{0}$
on the plaquette rungs increase up to their maximum possible value
of 1/2 at $J_2\approx 0.25$, while the inter-plaquette coupling amplitudes jointly decrease
to zero. Qualitatively similar behavior applies to $J_{1}$ values
other than that chosen in Fig.~\ref{fig5} within the red phase and
within the green phase by interchanging $J_{1}\text{\ensuremath{\leftrightarrow}}J_{2}$
and $A_{2},B_{1}\leftrightarrow B_{2},A_{1}$.

However, as signaled by the grey phases in Fig.~\ref{fig4}, and from
Fig.~\ref{fig5}, the SBMFT overestimates the stability of decoupled
singlet sub-units within the FFST - such as the four-spin-plaquette.
These grey phases are artifacts of the SBMFT which are reached through
second order transitions. As Fig.~\ref{fig5} shows, SBMFT allows
for small but finite parameter ranges with only one non-zero and maximized
AFM bond mean field, implying that the FFST decomposes into a collection
of completely decoupled $J_{0}$-, $J_{1}$-, or $J_{2}$-chains.
I.e.\ in the grey regions the SBMFT is incapable to lower the system
energy by quantum fluctuations between the latter decoupled chains. This is
the reason for the poor SBMFT ground state energy in Fig.~\ref{fig:egs}(b).

To conclude, also the SBMFT phase diagram is consistent with a gapped
plaquette phase extending throughout two strips of width of order unity
parallel to each of the $J_{1(2)}$-axis, and at least up to intermediate
$J_{1(2)}$. Within this plaquette phase $(\pi,\pi)((0,\pi))$-correlations
increase, as $J_{1}(J_{2})$ increase. Moreover SBMFT shows a $(\pi,0)$-phase,
similar to the classical case, however with a spin gap and without
long range order. We return to the latter in subsection \ref{LWC}.
Finally, SBMFT shows no incommensurate phase.


\subsubsection{Correlation functions and static structure factor\label{CfSf}}

In this section we turn to the question of a potentially incommensurate
phase in the quantum case. To this end we first look at static real-space
correlation functions \begin{equation}
C(\mathbf{r})=\langle\mathbf{S}(\mathbf{r})\cdot\mathbf{S}(\mathbf{0})\rangle,\label{eq:Corr}\end{equation}
 where $\mathbf{r}$ is a site on the lattice and $\langle\ldots\rangle$
the ground-state expectation value. Due to the $SU(2)$ invariance
of the model, only the correlation function $C_{z}(\mathbf{r})=\langle S_{z}(\mathbf{r})S_{z}(\mathbf{0})\rangle$
needs to be considered, which satisfies $C_{z}(\mathbf{r})=C(\mathbf{r})/3$.
We will contrast results from DMRG against those from SBMFT.

\begin{figure}[tb]
\begin{centering}
\includegraphics[width=0.3\columnwidth]{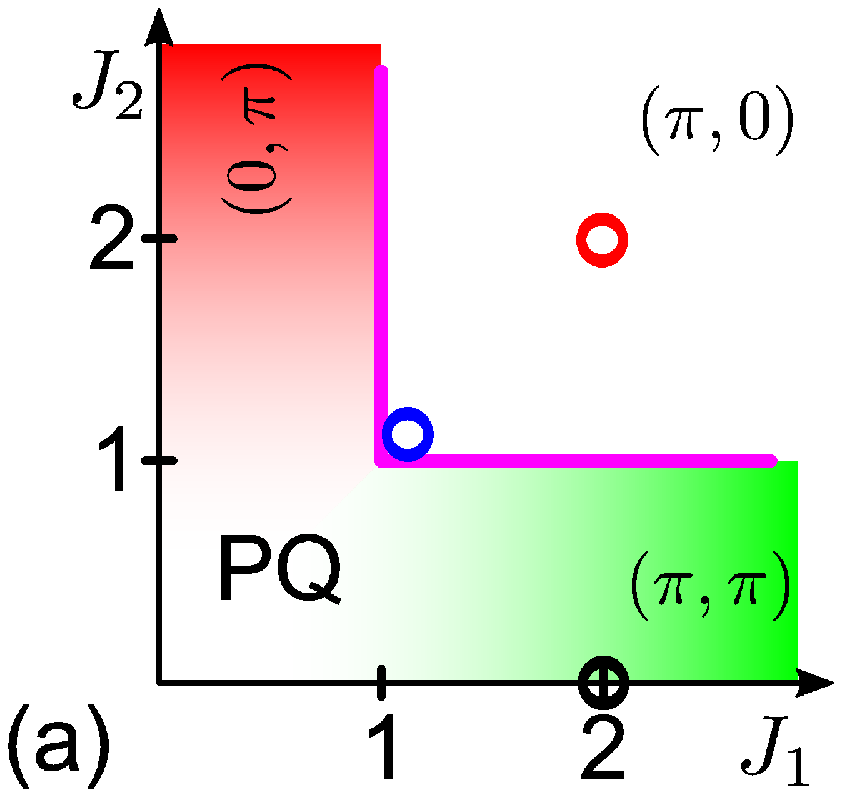}
\par\end{centering}

\begin{centering}
\includegraphics[width=0.8\columnwidth]{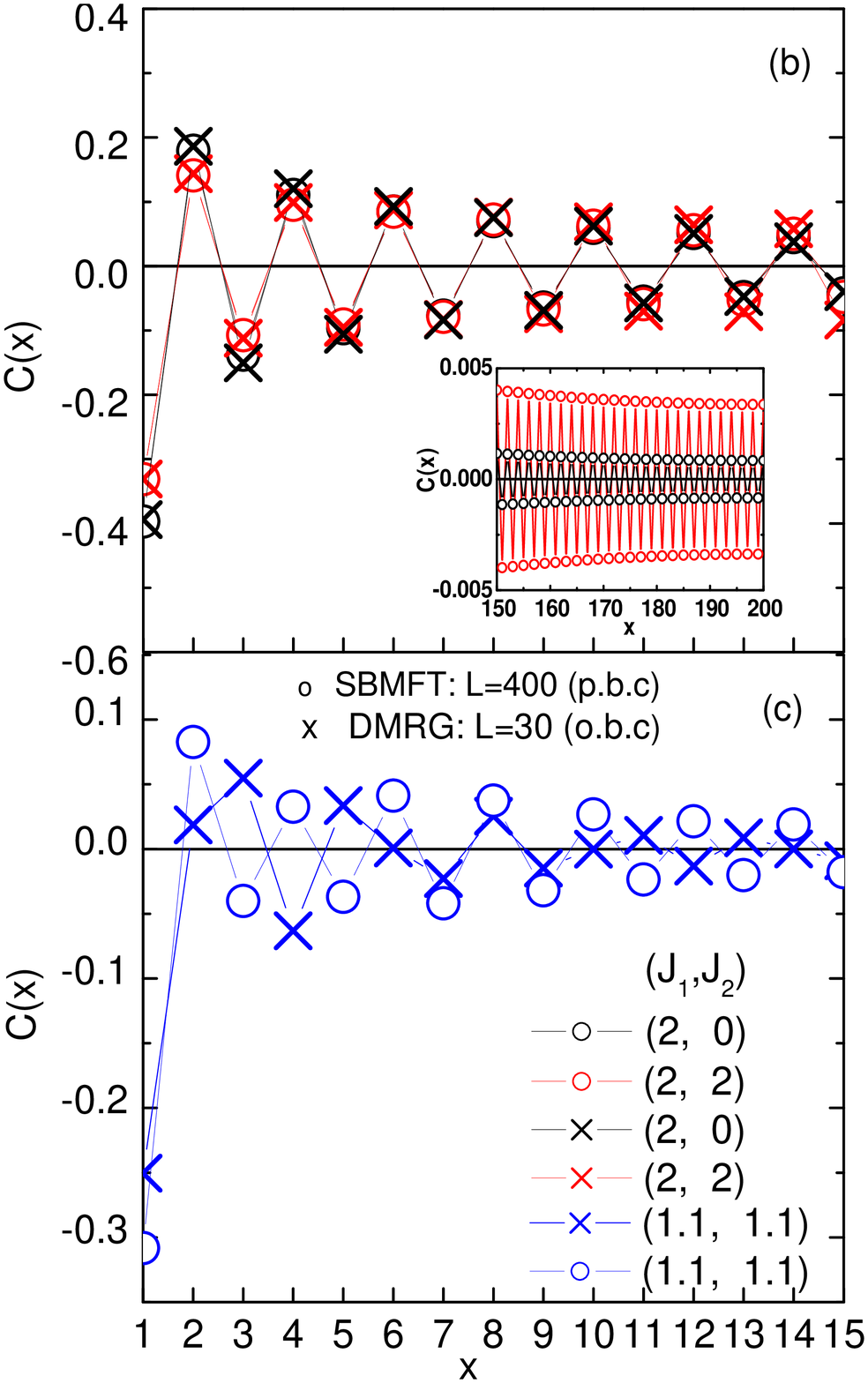}
\par\end{centering}

\caption{\label{corrSzSz} (color online) (a) Open colored circles: $J_{1,2}$ values chosen for
panels (b), (c). PQ refers to weakly coupled plaquette phase, with predominant
$(\text{\ensuremath{\pi}},\pi$) (green gradient) or $(0,\pi)$ (red
gradient) correlations. Magenta line: first order transition evidenced
from DMRG in Fig.~\ref{fig:egs}(b, inset)
and SBMFT in Fig.~\ref{fig4}.
(b) and (c) $C(x)=\langle\textbf{S}(x,1)\cdot\textbf{S}(L/2,1)\rangle$
vs $x$. Curve colors correspond to choices in (a). Crosses: DMRG
for L=30, m=300, and OBC. Circles: SBMFT for L=400 and PBC. Inset
in (b): SBMFT at large distance.}
\end{figure}

SBMFT results are obtained with periodic boundary conditions (PBC).
For best
convergence, DMRG employs open boundary conditions (OBC) along the
chain. I.e., correlations depend on the reference site. To minimize
edge effects, we have chosen a reference site $\mathbf{0}=(L/2,y)$ in the
middle of any of the $y=(1\ldots4)$ equivalent chains of the tube.
Fig. \ref{corrSzSz}, panels(b), (c), shows
$C(x)$
along one of those equivalent chains,
say  $\mathbf{r}=(L/2-1+x,1)$ and $\mathbf{0}=(L/2,1)$.

We have focused on three particular
values of $J_{1,2}$ as shown in the schematic phase diagram in panel (a).
Two of them lie regions where both,
the classical and the SBMFT suggest strongly commensurate correlations,
and one is shortly \emph{above} the first order transition of Fig.~\ref{fig:egs},
where the classical state is incommensurate.

Fig. \ref{corrSzSz}(b) evidences clearly commensurate correlations
along the tube's legs for the regions of the black and red open circles
in Fig.~\ref{corrSzSz}(a) and obviously a remarkably good agreement
between DMRG and SBMFT. Small deviations between DMRG and SBMFT at
the ends of the chain are to be expected from the difference in boundary conditions.
We have checked, that the
wave vector of the commensuration is $(\pi,\pi)$($(\pi,0)$) for
the black(red) circles of \ref{corrSzSz}(a) by also scanning along
other real-space directions on the FFST. Clearly $C(x)$ decays as
a function of $x$. While the system sizes for the DMRG are too small
to extract the functional form of this decay, $C(x)\sim\exp(-x/\xi)$
is found in the SBMFT, where $\xi$ is a finite correlation length
related with the inverse of the energy gap.
This is consistent with gapped phases and no LRO, as
has already been alluded to in section \ref{sub:SBMFTphadig}.

The situation changes drastically at the blue open circle in fig.
\ref{corrSzSz}(a). Here, DMRG evidences a strongly decaying, \emph{incommensurate}
$x$-dependence in Fig.~\ref{corrSzSz}(c), while SBMFT continues
to display commensurate $(\pi,0)$-correlations, as to be expected
from the phase diagram, Fig.~\ref{fig4}. This proves, that SBMFT
fails to produce the proper spin-correlations shortly above the first
order transition out of the plaquette phase and suggests the presence
of an incommensurate region also in the quantum version of the FFST.

\begin{figure}[tb]
\begin{centering}
\includegraphics[width=0.35\columnwidth]{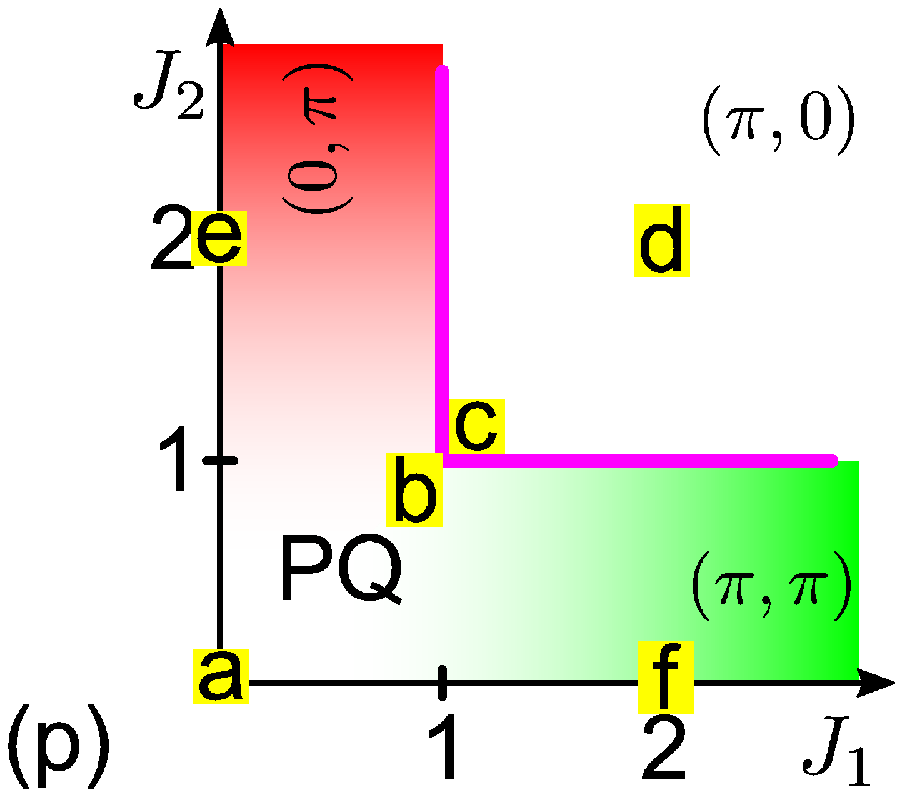}
\par\end{centering}

\begin{centering}
\includegraphics[width=8.5cm]{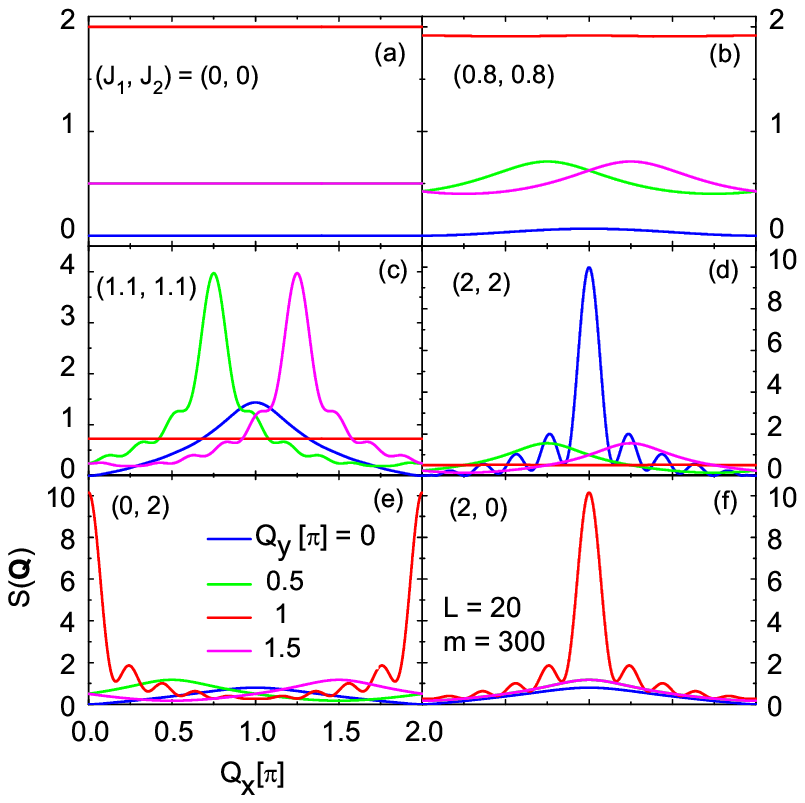}
\par\end{centering}

\caption{\label{fig:Sf-cuts} (color online) (p) Letters on yellow background: $J_{1,2}$
choices for panels (a)-(f). PQ refers to weakly coupled plaquette phase,
with predominant $(\text{\ensuremath{\pi}},\pi$) (green gradient)
or $(0,\pi)$ (red gradient) correlations. Magenta line: first order
transition evidenced from DMRG, Fig.~\ref{fig:egs}(b, inset)
and SBMFT Fig.~\ref{fig4}. Panels (a)-(f): structure factor $S(\textbf{Q})$
from DMRG (L=20, m=300,
OBC) for $J_{1,2}$ as in (p). Blue, green, red, and magenta lines
refer to $Q_{y}=(0,1,2,3)\pi/2$.}
\end{figure}

To further corroborate this, we now calculate the static structure
factor
\begin{equation}
S(\mathbf{Q})=\frac{1}{4L}\sum_{\mathbf{r}}e^{i\mathbf{Q}\cdot\mathbf{r}}\langle\mathbf{S}(\mathbf{r})\cdot\mathbf{S}(\mathbf{0})\rangle,\label{eq:Sf}
\end{equation}
versus wave vector $\mathbf{Q}=(Q_{x},Q_{y})$ from our DMRG data where $\mathbf{0}=(L/2,1)$.
First we consider a coarse grained set of $J_{1,2}$. The results
are shown in Fig.~\ref{fig:Sf-cuts}. As labeled in panel (a) four
values of $J_{1,2}$ are taken from regions where commensurate correlations
are to be expected and two out of the vicinity of the first order
transition as observed in DMRG, Fig.~\ref{fig:egs}(b, inset)
and SBMFT Fig.~\ref{fig4}. Since the transverse momentum space of
the tube is confined to $Q_{y}=(0,1,2,3)\pi/2$ there are four $S(Q_{x},Q_{y})$-lines
for each value of $J_{1,2}$.

Fig. \ref{fig:Sf-cuts}(a) exhibits a flat structure for all $Q_{y}$
modes vs.~$Q_{x}$, which reflects the decoupling of the plaquettes.
Moreover $S(\mathbf{Q})$ is maximum at $Q_{y,max}=\pi$ consistent with
the singlet ground state on the decoupled plaquettes. Figs. \ref{fig:Sf-cuts}(d,e,f)
show maxima in $S(\mathbf{Q})$ at $\mathbf{Q}_{max}=(\pi,0)$, $(0,\pi)$,
and $(\pi,\pi)$ respectively. This is consistent with SBMFT in Fig.~\ref{fig4}
and also with the classical phase diagram in Fig.~\ref{fig2}. The
small oscillations around the maxima are finite size effects. On the
\emph{finite} system used for the DMRG calculations, the amplitude
of the structure factor remains finite at $\mathbf{Q}_{max}$. From
the analysis up to now, we expect no LRO on the quantum FFST, i.e.\
a finite value of $S(\mathbf{Q}_{max})$ for $L\rightarrow\infty$.
A proof of the latter would require finite size scaling analysis,
which is beyond our computational reach.

Figs. \ref{fig:Sf-cuts}(b,c) describe $J_{1,2}$ values shortly below and above the first
order transition of Fig.~\ref{fig:egs}(b, inset),
along the line of maximum frustration. Panel (b) contains a small
modulation in all modes, although the plaquette phase is still evident
from $Q_{y,max}=\pi$. Panel (c) however shows two-symmetric
maxima at \emph{incommensurate} vectors with
$ \mathbf{Q}_{max}=(3\pi/4,\pi/2),\,(5\pi/4,3\pi/2)  $.
While the $y$-component of these pitch
vectors are set by the transverse quantization of the momentum space,
the $x$-components are set by the quantum correlations in the FFST.
Very remarkably, these $x$-components are, up to our numerical precision ($10^{-4}$), identical to the
corresponding classical pitch-vectors, listed in the enumeration point
\ref{enu:Two-degenerate-incommensurate} in section \ref{sec:Classical-phase-diagram}.

\begin{figure}[tb]
\begin{centering}
\includegraphics[width=0.9\columnwidth]{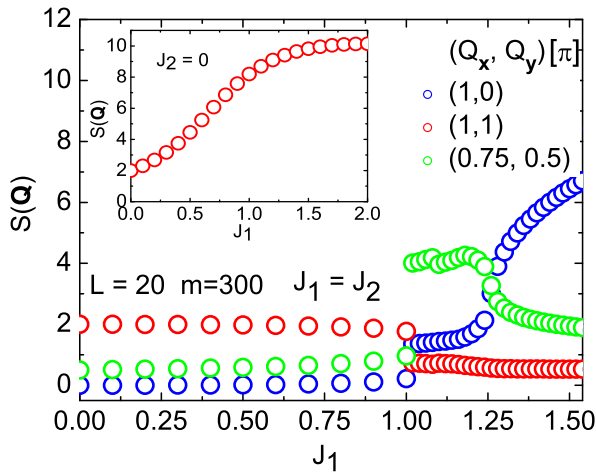}
\par\end{centering}

\caption{\label{fig:Sf-q} (color online) $S(\mathbf{Q})$ from DMRG, L=30, m=300, and
OBC versus $J_{1}=J_{2}$ at $\mathbf{Q}=(\pi,0)$ (blue circles),
$(\pi,\pi)$ (red circles) and for the incommensurate
$\mathbf{Q}_{max}=\pm(3\pi/4,\pi/2)$ (green circles).
Inset: $S(\pi,\pi)$ versus $J_{1}$ at $J_{2}=0$.}
\end{figure}

Next, we discuss the DMRG structure factor in a finer grained analysis of the $J_{1,2}$ plane,
along the diagonal line of maximum frustration.
We have observed the occurrence of $Q_{y,max}=\pi$ with flat $Q_x$ dependence,
characteristic of the plaquette phase, for $J_1=J_2 \lesssim 0.7$ and
$\mathbf{Q}_{max}=(\pi,\pi))$, still with very flat $Q_x$ dependence,  for $0.8 \lesssim J_1=J_2\lesssim 1$
(cf.\ Figs.~\ref{fig:egs}(a,b)).
We also find  $\mathbf{Q}_{max}= (\pi,0)$, signaling a commensurate classical-like $(\pi,0)$ phase for $J_1=J_2\gtrsim 1.3$
(cf.\ Fig.~\ref{fig:egs}(d)).
An incommensurate phase is observed for $1 \lesssim J_1=J_2 \lesssim 1.3$, with $ \mathbf{Q}_{max}= (3\pi/4,\pi/2),\,(5\pi/4,3\pi/2)$.

In order to describe the extent of such incommensurate region
we show in Fig.~\ref{fig:Sf-q} $S(\mathbf{Q})$ for representative momenta
$\mathbf{Q}=(\pi,\pi),\,  (3\pi/4,\pi/2), (\pi,0) $ in the range $J_1=J_2\in [0,1.5]$.
Clearly, $S(\pi,\pi)$ is maximum and shows only a small
variation in the range $J_{1}=J_{2}\in[0,1]$ (plaquette phase).
At $J_{1}=J_{2}=1$, the structure factor is discontinuous.
Following that, and in a small window of $1\lesssim J_{1}=J_{2}\lesssim1.2$,
$S(\mathbf{Q})$ is maximum at the incommensurate wave vector. In
the vicinity of $J_{1}=J_{2}\approx1.25$ there is a crossover from
incommensurate to commensurate $(\pi,0)$ correlations. These results
can be interpreted in terms of a small window of an incommensurate
phase with a weak first order transition into the $(\pi,0)$-phase
and a kink in the energy which is too small to be detected
from the DMRG calculations in Fig.~\ref{fig:egs}.

For reference the inset in Fig.~\ref{fig:Sf-q} reports $S(\pi,\pi)$
along the $J_{1}$-axis, i.e.\ $J_{2}=0$, where the structure factor is maximum
for any $J_{1}>0$.
This plot shows a continuous
increase and no signs of phase transitions in this part of parameters
space. An identical observation applies to $S(0,\pi)$ along
the $J_{2}$-axis, i.e.\ $J_{1}=0$, for all $J_{2}>0$. This is consistent
with the plaquette phase being adiabatically connected with the limit
of decoupled chains.

While the discussion in Fig.~\ref{fig:Sf-q} is confined
to the line of maximum frustration, we have performed similar analysis
along additional lines in the $J_{1,2}$ plane. These agree with a
plaquette phase in strips of width one, both, along the $J_{1}$-, and
$J_{2}$-axis, as in Fig.~\ref{fig4}, up to values of $J_{1,2}\approx 5$.
This extends the range obtained from the kink in the ground
state energy in Fig.~\ref{fig:egs} and Ref.~\onlinecite{Arlego2011}.
Moreover, incommensurate correlations are observed beyond these strips,
with  $\mathbf{Q}_{max}$ slightly renormalized by quantum fluctuations
with respect to the classical spiral pitch-vectors in the enumeration point
\ref{enu:Two-degenerate-incommensurate} in section \ref{sec:Classical-phase-diagram}.
Unfortunately, the width of the incommensurate region decreases rapidly off from
the line of maximum frustration and cannot be determined accurately
enough.

To summarize, static structure factor calculations suggest that the
at least close to line of maximum frustration, the plaquette phase-strips
undergo a first order transition into an incommensurate phase, the
extent of which is strongly decreased by quantum fluctuations with
respect to the classical spiral phase. The transition between the incommensurate
and the $(\pi,0)$ phase appears to be very weakly first order.


\subsection{Strong leg coupling\label{LWC} }

In this subsection we consider the limit of $J_{1,2}\gg J_{0}$ by
explicitly setting $J_{0}=0$. Naturally in this limit, a different
normalization is required for the exchange coupling constants. We
set $J_{1}=1$. For $J_{0}=0$ the FFST is unfrustrated,
equivalent to an anisotropic twisted square lattice on a torus.
We will use quantum Monte-Carlo, based on the stochastic series expansion,
to study its properties.


\subsubsection{Uniform susceptibility and spin gap}

The real space arrangement of spins in the classical $(\pi,0)$ phase
at $J_{1}\sim J_{2}\gg J_{0}$ is that of a spin-$2$ AFM chain. While
in the quantum model the total spin per plaquette is not conserved,
it is nevertheless tempting to speculate on a gap similar to that
of an actual spin-2 AFM quantum chain at $J_{2}=J_{1}$. Additionally,
upon reducing $J_{2}/J_{1}\text{\ensuremath{\rightarrow}}0$, the
limit of four decoupled chains is reached, which is a LLQ and therefore
shows no spin gap.

\begin{figure}[tb]
\begin{centering}
\includegraphics[width=0.8\columnwidth]{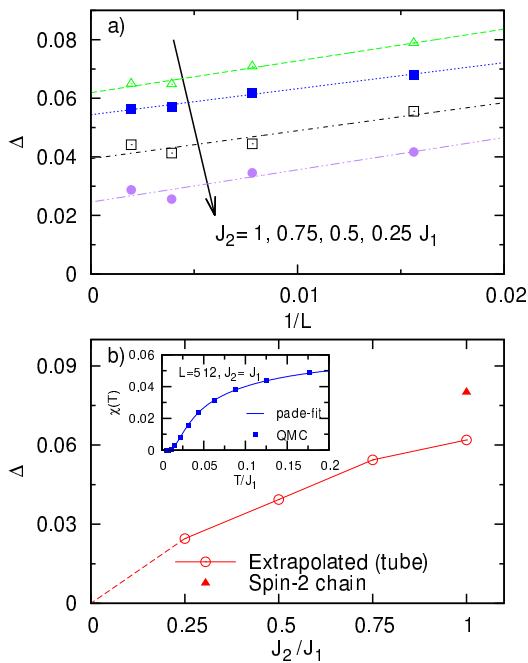}
\par\end{centering}

\caption{\label{QMCfig1} (color online) a) Finite size scaling for $64\leqslant L\leqslant512$
of the spin gap for different values of $J_{2}$. b) $L=\infty$ extrapolated
spin gap vs. $J_{2}$, as well as spin gap of a spin-2 chain. Inset:
susceptibility vs. temperature for $L=512$ and $J_{2}=J_{1}$. Symbols:
QMC data, solid curve: Padé fit, see text.}
\end{figure}

To test these assumptions, we evaluate the uniform spin susceptibility
$\chi(T)$ versus temperature $T$ on systems of up to $L=512$ plaquettes
for $J_{2}=1,\,0.75,\,0.5$, and $0.25J_{1}$. The case of $J_{2}=J_{1}$
is shown in the inset of Fig.~\ref{QMCfig1}(b). Obviously, the system
has gap. The value of the gap is extracted from $\chi(T)$ by fitting
the low-temperature behavior for $0.0055\leqslant T\leqslant0.2J_{1}$
to $\chi(T)\approx e^{-\Delta/T}P_{k}^{l}(T)/T$, where $P_{k}^{l}(T)$
is a Padé approximant of order $[k,l]$. The errors of such fits -
for a particular choice of the fitted temperature interval - can be
made less than the QMC's error bars, which are not shown in Fig.~\ref{QMCfig1},
and are of order of $10^{-6}$. Fig. \ref{QMCfig1}(a) details the
finite size scaling of the spin gap for $64\leqslant L\leqslant512$.
The small oscillations of the data in this plot should not be confused
with QMC errors or deviations from simple scaling. Rather they are
due to the particular choice of the temperature interval for the Padé
fit. As is obvious from this figure, these oscillations are less than
the actual finite size corrections. Finally, the main panel of fig.
\ref{QMCfig1}(b) proves our speculation, namely, the spin gap at
$J_{2}=J_{1}$ is close to that of a spin-2 chain \cite{Grossjohann2010}
and the gap decreases monotonously as $J_{2}/J_{1}\text{\ensuremath{\rightarrow}}0$,
where, corresponding to the LLQ,
$\Delta(J_{2}/J_{1}=0)=0$.


\subsubsection{Dynamic structure factor}

\begin{figure}[t]
\begin{centering}
\includegraphics[width=1\columnwidth]{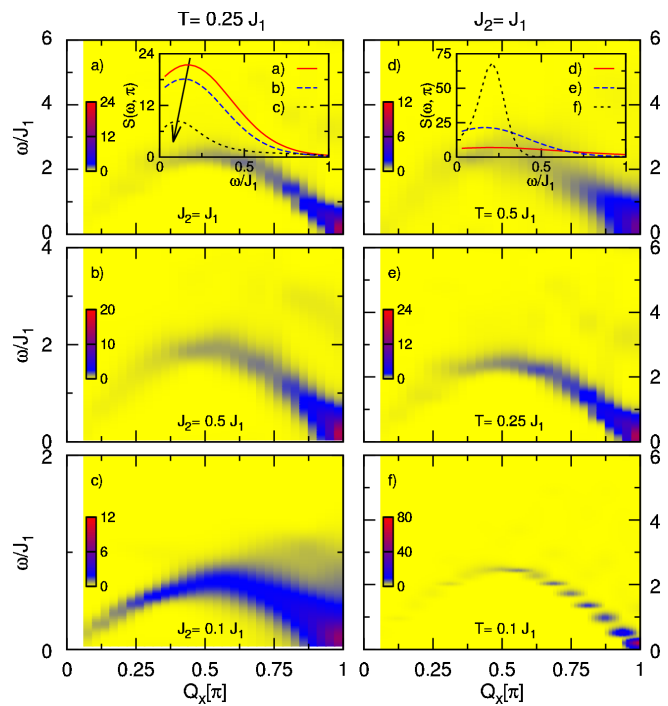}
\par\end{centering}

\caption{\label{QMCfig2} (color online) Contour plots of the dynamic structure factor $S(\mathbf{Q},\omega)$
from QMC \& MaxEnt for systems of $L=32$ vs. $\omega$ and $Q_{x}$
at $Q_{y}=0$. Panels (a-c): for $J_{2}/J_{1}=1$, $0.5$, and $0.1$
at $T=0.25J_{2}$. Panel a) inset: $(\pi,0)$-cut of $S(\mathbf{Q},\omega)$
with labels referring to panels a)-c). Panels (d-f): for $T/J_{1}=0.5$,
$0.25$, and $0.1$ at $J_{2}/J_{1}=1$. Panel d) inset: $(\pi,0)$-cut
of $S(\mathbf{Q},\omega)$ with labels referring to panels d)-f).}
\end{figure}

Continuing on the analogy of a crossover from a gapped Haldane-like
spin-2 AFM chain to a LLQ for $J_{2}/J_{1}$ ranging from $1$ to
$0$, the dynamical structure factor of the FFST should show signatures
of deconfinement from gapped 'magnon'-like modes at $J_{2}/J_{1}=1$
to a two-spinon continuum as $J_{2}/J_{1}\text{\ensuremath{\rightarrow}}0$.

To analyze this, we investigate the dynamic structure factor $S(\mathbf{Q},\omega)$
at frequency $\omega$, which we obtain from MaxEnt analytic continuation
of imaginary time dynamic structure factor
\[
S(\mathbf{Q},\tau)=\frac{1}{4L}\sum_{\mathbf{r}}e^{i\mathbf{Q}\cdot\mathbf{r}}\langle\mathbf{S}(\mathbf{r},\tau)\cdot\mathbf{S}(\mathbf{0},0)\rangle,
\]
evaluated by QMC (see appendix \ref{QMC-theory}). This is shown in
Fig.~\ref{QMCfig2}. In all of these plots $Q_{y}=0$. The absolute
scales on all panels of this figure are adjusted to ensure approximately
identical extent of the spectra along the $y$-axes, which allows
to compare the width of the spectral contours. Turning to Fig.~\ref{QMCfig2}(a-c),
we first note that all three contour plots display a certain broadening
due to the finite temperature $T=0.25J_{1}$. We return to this in
Fig.~\ref{QMCfig2}(d-f). Apart from that, at $J_{1}=J_{2}$ the figures
show a rather sharp magnon-like mode, similar to the spectra of integer-spin
Haldane chains, \cite{Meshkov1993,Grossjohann2010} accompanied by
a marked loss of spectral weight as $Q_{x}\rightarrow0$, which is
also a typical feature of integer spin chains. \cite{Ma1992} As $J_{2}\rightarrow0$,
the spectrum starts to broaden in the vicinity of $Q_{x}=\pi$, resembling
a shape very similar to that of the spinon continuum of the spin-1/2
AFM Heisenberg chain \cite{Grossjohann2009,Caux2011} - exactly as
anticipated. The inset in Fig.~\ref{QMCfig2}(a) details, that although
the finite temperature maximum of the dynamic structure factor does
not have to coincide with the spin gap, it nevertheless decreases
similar to the latter with respect to $J_{2}/J_{1}$.

Figs. \ref{QMCfig2}(d-f) list the temperature dependence of
$S(\mathbf{Q},\omega)$ for $J_{2}=J_{1}$. First, these panels clarify, that
$T=0.25J_{1}$ is a reasonable compromise between finite size effects at $L=32$
and thermal broadening, i.e., for $T=0.1J_{1}$ the line broadening is already
less than the finite-size level-spacing. Furthermore, the inset \ref{QMCfig2}(d)
collects cuts at $Q_{x}=\pi$, which demonstrate a rather strong temperature
dependence of the zone-boundary modes of the FFST for $J_{2}\approx J_{1}\gg
J_{0}$. This might be of interest in the context of similar observations
\cite{Zheludev2008a} for four-spin tube compound
Cu$_{2}$Cl$_{4}$$\cdot$D$_{8}$C$_{4}$SO$_{2}$.

In conclusion, and even though the on-plaquette total spin is not strictly
conserved in the quantum case, the FFST shows some features remarkably similar
to an AFM chain with $S\approx 2$ at $J_1=J_2\gg J_0$, as well as a
magnon-spinon deconfinement as $J_2/J_1\rightarrow 0$.


\section{Discussion and Conclusions\label{sec:Discussion-and-Conclusions}}

\begin{figure}[tb]
\begin{centering}
\includegraphics[width=0.75\columnwidth]{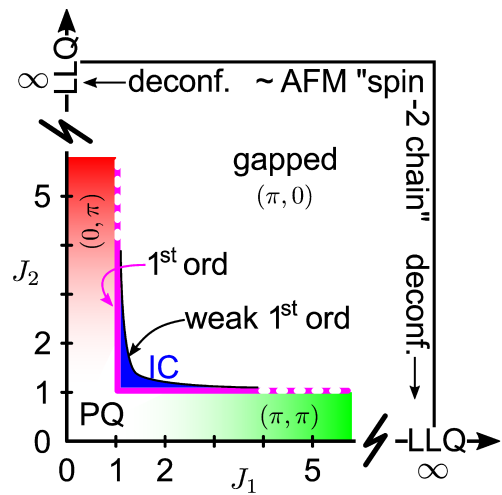}
\caption{\label{fig11} (color online) Quantum phase diagram of the FFST}
\par\end{centering}
\end{figure}

To summarize, we have studied the quantum phases of a frustrated spin-1/2
four-leg tube using a variety of techniques: density
matrix renormalization group, Quantum Monte-Carlo, Schwinger boson mean field
theory, exact diagonalization and series expansions.  Our main results are
outlined in the tentative quantum phase diagram Fig.~\ref{fig11}. This
figure should be contrasted against the tube's phase diagram in the classical
limit, i.e. Fig.~\ref{fig2}. While all phases in the latter are long range ordered,
none of the quantum phases are.

The point $J_{1,2}=0$ hosts a gapped system of decoupled plaquettes, while at the
asymptotic points $J_{1(2)} \to\infty$, $J_{2(1)} = 0$ the spin tube degenerates
into decoupled spin-1/2 chains in a Luttinger liquid state. The phase diagram is
symmetric with respect to interchanging $J_1 \leftrightarrow J_2$. On either of
the two axes $J_{1(2)} =0$, the system is unfrustrated, the inter-leg coupling
is relevant, and a spin gap opens. This unfrustrated weakly coupled chain
regime is known to be adiabatically connected to that of the weakly coupled
plaquettes.

Turning on the frustrating exchange, our results are consistent with the weakly
coupled plaquette regime to survive along two strips (red and green in
Fig.~\ref{fig11}) of width of order unity, parallel to each of the
$J_{1(2)}$-axis, at least up to $J_{1(2)} \approx 5 J_0$.  The system remains
gapped in this region. Accordingly, our analysis of correlation functions
exhibits exponential real space decay. Consistent with series expansions around
$J_{1,2}=0$, the static structure factor obtained from density matrix
renormalization group evolves smoothly from a flat plaquette signature around PQ
in Fig.~\ref{fig11}, into a peaked commensurate behavior along the red/green
strips, parallel to each axis. The peak locations are consistent with
short-range correlation remnants of the long-range order present in the
classical limit of the tube this region. As for the unfrustrated four-leg tube,
we expect no quantum phase transition while increasing
$J_{1(2)}\rightarrow\infty$ parallel to the axis within these strips until the
Luttinger liquid fixed point is reached (zig-zag marks in Fig.~\ref{fig11}).

Perpendicular to the $J_{1,2}$-axis the plaquette regime is terminated by a line
of first order transitions evidenced by those of our techniques able to detect
ground state energy level crossings. The critical lines emerge approximately
from the point of maximum frustration $J_{0,1,2}=1$ and run parallel to the
$J_{1,2}$ axes (magenta line in Fig.~\ref{fig11}). The numerical
precision, locating the level crossing along the borders of the PQ strip,
decreases away from $J_{0,1,2}=1$, indicated by the doting of the magenta line.

Beyond the first order critical line, close to the point of maximum frustration,
$J_{0,1,2}=1$, DMRG shows that the plaquette phase turns into a gapped phase
with short range incommensurate correlations (IC, blue in Fig.~\ref{fig11}),
analogous to the spiral phase which is found in the classical limit of the tube
in this regime. Along the diagonal $1 \lesssim J_1=J_2 \lesssim 1.3$, the static
structure factor shows a maximum approximately at the pitch vectors of the
classical spiral phase.  Off the diagonal, the maximum of the static structure
factor is slightly shifted from the classical values. Increasing the
inter-plaquette coupling, around the line $J_1\sim J_2$, the incommensurate
quantum phase terminates with a very weak first order transition into a gapped
commensurate ($\pi$, 0) phase, labeled by the thin black line in
Fig.~\ref{fig11}. In contrast to the PQ $(\pi,\pi)$ and $(0,\pi)$ region, the
overall extent of the incommensurate region in the quantum case is strongly
reduced as compared to that of the classical spiral phase.

For $J_{1,2}\gg J_0$, the system can be considered as approximately
unfrustrated. We have investigated this regime by Quantum Monte-Carlo along the
line $0<J_2/J_1\leq 1$, setting $J_0=0$ (black line emerging perpendicularly
from the point LLQ in Fig.~\ref{fig11}). Here calculations of the uniform
susceptibility show the tube to have a gap very close to that of AFM spin-2
chains at $J_1=J_2$, while for $J_2/J_1\rightarrow 0$ the gap decreases to zero
as expected for approaching the Luttinger liquid state. Evaluating the dynamic
structure factor, and consistent with a crossover from a 'Haldane-like AFM
spin-2 chain' behavior at $J_1\sim J_2$ to a LLQ at $J_2=0$, we observe a
deconfinement of the excitations turning from sharp 'magnon' modes into a spinon
continuum as $J_2/J_1\rightarrow 0$.

Finally, and due to numerical limitations in our study, it remains an open issue
if the quantum IC and PQ regimes extend beyond $J_{1(2)}\sim 5$ at $J_{2(1)}\sim
1$. In this context we cannot conclude from our study whether the PQ and
$(\pi,0)$ phase remain adiabatically disconnected in the quantum case or not.

\section{Acknowledgments}

We thank D.C.~Cabra for helpful discussions. MA, HDR and GR have
been supported by CONICET (PIP 1691) and by ANPCyT (PICT 1426). Part
of this work has been supported by the Deutsche Forschungsgemeinschaft
through FOR912, Grant No. BR 1084/6-2 (WB), the European Commission
through MC-ITN LOTHERM, Grant No. PITN-GA-2009-238475 (YR and WB),
and the NTH School for Contacts in Nanosystems (BW and WB). WB thanks
the Platform for Superconductivity and Magnetism Dresden, and the
Kavli Institute for Theoretical Physics for kind hospitality. The
research at KITP was supported by the National Science Foundation
under Grant No. NSF PHY11-25915.

\appendix

\section{Techniques\label{sec:Techniques}}

For completeness, this appendix provides some details and references
to the methods we use in this work.

\subsection{Series expansion\label{sub:Series-expansion}}

Our SE calculations start from the limit of isolated plaquettes. To
this end we decompose the Hamiltonian of the FFST into \begin{equation}
H=H_{0}+V(J_{1},J_{2}),\label{se1}\end{equation}
where $H_{0}$ represents decoupled plaquettes and $V(J_{1},J_{2})$
is the part of Hamiltonian that connects plaquettes via $J_{1},J_{2}$
couplings.

It is simple to show, that each plaquette has four equally spaced energy
levels, which in turn renders the levels structure of $H_{0}$ to
be equidistant. This allows to sort the spectrum of $H_{0}$ in a
block-diagonal form, where each block is labeled by an energy quantum-number
Q. In this way, Q=0 represents the ground state (\emph{vacuum}), i.e.,
all plaquettes are in the state of minimum energy. Q=1 sector is composed
by states obtained by creating (from vacuum state) one-elementary
excitation (\emph{particle}) on a given plaquette, and so on. It is
clear that $Q\geq2$ will be of multiparticle nature.

In general the action of $V(J_{1},J_{2})$ mixes different Q-sectors,
so that the block-diagonal form of $H_{0}$ is not conserved in $H$.
However, it has been shown \cite{Knetter2000a} that for the present
type of Hamiltonians it is possible to restore block-diagonal form
by the application of continuous unitary transformations, using the
flow equation method of Wegner \cite{Wegner1994a}. It basically consists
in transforming $H$ onto an effective Hamiltonian $H_{\mathrm{eff}}$
which is block-diagonal in the quantum number $Q$. This transformation
can be achieved exactly in terms of a SE in $J_{1,2}$ leading to
\begin{equation}
H_{\mathrm{eff}}=H_{0}+\sum_{n,0\leq m\leq n}J_{1}^{n-m}J_{2}^{m}C_{n,m}\,.\label{eq:5}\end{equation}
Here $C_{n,m}$ are weighted products of terms in $V(J_{1},J_{2})$
which conserve the $Q$-number, with weights determined by recursive
differential equations (see Ref. \cite{Knetter2000a} for details).

Due to $Q$-number conservation several observables can be calculated
directly from $H_{\mathrm{eff}}$ in terms of a SE in $J_{1,2}$.
For systems with coupled spin-plaquettes continuous unitary transformations
SE has been used for one \cite{Arlego2006a}, two \cite{Arlego2008a,Arlego2007a,Brenig2004a,Brenig2002aa}
and three \cite{Brenig2003a} dimensions. For the present model we
have performed $O(7)$ and $O(6)$ SE in $J_{1,2}$ for ground state
energy ($Q=0$) and for $Q=1,2$ sectors, respectively. We refer for
technical details about the calculation to Ref. \cite{Arlego2011}.

\subsection{Schwinger bosons\label{sub:Schwinger-bosons}}

Schwinger bosons \cite{Auerbach} are used to represent spins at site
$l$ via spinfull bosons $b_{l\sigma}^{(\dagger)}$, with $\sigma=\uparrow\downarrow$
or $\pm1$, through $S_{l}^{\alpha}=\frac{1}{2}\sum_{\mu\nu}b_{l\mu}^{\dagger}\sigma_{\mu\nu}^{\alpha}b_{l\nu}^{\phantom{\dagger}}$,
where $\sigma_{\mu\nu}^{\alpha}$ are the Pauli matrices and $\alpha=x,y,z$.
The Hilbert space dimension of spin-$S$ multiplets is enforced through
the constraint $\sum_{\sigma}b_{l\sigma}^{\dagger}b_{l\sigma}^{\phantom{\dagger}}$=$2S$.
In terms of Schwinger bosons, the exchange interaction can be written
as \cite{AA,Auerbach}
\begin{equation}
\mathbf{S}_{l}\cdot\mathbf{S}_{m}=\,:\hat{B}_{lm}^{\dagger}\hat{B}_{lm}^{\phantom{\dagger}}:-\hat{A}_{lm}^{\dagger}\hat{A}_{lm}^{\phantom{\dagger}}\,,\label{w1}\end{equation}
with the bond operators $\hat{B}_{lm}^{\dagger}$ = $\frac{1}{2}\sum_{\sigma}b_{l\sigma}^{\dagger}b_{m\sigma}^{\phantom{\dagger}}$
and $\hat{A}_{lm}$ = $\frac{1}{2}\sum_{\sigma}\sigma b_{l\sigma}b_{m-\sigma}$
and normal ordering $::$. Eqn. (\ref{w1}) has been used for various
$SU(2)$ invariant and large $N$ factorization schemes \cite{AA,Read1991,Sachdev1992,TGC,Flint2009}.
We follow \cite{TGC,Flint2009} and introduce the\emph{ }bond mean
fields\emph{ $B_{lm}=\langle\hat{B}_{lm}\rangle$} and $A_{lm}=\langle\hat{A}_{lm}\rangle$,
accounting for ferromagnetic (FM) and AFM correlations on equal footing.
For the FFST we focus on homogeneous mean fields, implying \emph{six
}parameters \begin{equation}
B_{n=0,1,2}\hphantom{aaaa}A_{n=0,1,2}\,,\label{w0}\end{equation}
where $n=0,1,2$ corresponds to the three exchange links $\mathbf{r}_{l}-\mathbf{r}_{m}$=$\mathbf{r}_{n}$=$\mathbf{R}_{y}$,
$\mathbf{R}_{x}$, $\mathbf{R}_{x}+\mathbf{R}_{y}$. Fourier transformation,
$b_{l\sigma}=\sum_{\mathbf{k}}e^{i\mathbf{k}\cdot\mathbf{r}_{l}}b_{\mathbf{k}\sigma}/\sqrt{N}$,
leads to a bilinear mean field Hamiltonian, which can be diagonalized
by standard Bogoliubov transformation, i.e.\ $b_{\mathbf{k}\sigma}^{\phantom{\dagger}}=u_{\mathbf{k}}a_{\mathbf{k}\sigma}^{\phantom{\dagger}}+i\, v_{\mathbf{k}}a_{-\mathbf{k}-\sigma}^{\dagger}$,
with $u_{\mathbf{k}}^{2}-v_{\mathbf{k}}^{2}=1$ yielding \begin{eqnarray}
H_{\mathrm{MFT}} & = & \sum_{\mathbf{k}\sigma}E_{\mathbf{k}}\left(a_{\mathbf{k}\sigma}^{\dagger}a_{\mathbf{k}\sigma}^{\phantom{\dagger}}+\frac{1}{2}\right)+\sum_{n}J_{n}\left(\left|A_{n}\right|^{2}-\right.\nonumber \\
 &  & \left.\left|B_{n}\right|^{2}\right)+2N\lambda(S+\frac{1}{2})\,,\label{w2}\end{eqnarray}
where $E_{\mathbf{k}}=[\gamma_{B}(\mathbf{k})^{2}-\gamma_{A}(\mathbf{k})^{2}]^{1/2}$
is the quasiparticle dispersion with $\gamma_{A}(\mathbf{k})$ = $\sum_{n}J_{n}A_{n}\sin(\mathbf{k}\cdot\mathbf{r}_{n})$
and $\gamma_{B}(\mathbf{k})$ = $\sum_{n}J_{n}B_{n}\cos(\mathbf{k}\cdot\mathbf{r}_{n})-\lambda$.
We assume $B_{n},A{}_{n}$ to be real. $\lambda$ is a Lagrange parameter
to enforce the constraint on the \emph{average}. Selfconsistency,
i.e.\ $\partial\langle H_{\mathrm{MFT}}\rangle/\partial x$=0, with
$x$=$A_{n},B_{n}$, and $\lambda$ leads to \begin{eqnarray}
A[B]_{n} & = & \frac{1}{2N}\sum_{\mathbf{k}}\frac{\gamma_{A[B]}(\mathbf{k})\sin[\cos](\mathbf{k}\cdot\mathbf{r}_{n})}{E_{\mathbf{k}}}\label{w3}\\
(S+\frac{1}{2}) & = & \frac{1}{2N}\sum_{\mathbf{k}}\frac{\gamma_{B}(\mathbf{k})}{E_{\mathbf{k}}}\,,\label{w4}\end{eqnarray}
where eqn. (\ref{w3}) yields six equations for $A_{n}$ and $B_{n}$,
by replacing terms with their square bracketed successors.

To obtain $A_{n},B_{n}$, and $\lambda$ we use two numerical approaches:
(i) we solve eqn. (\ref{w3},\ref{w4}) in the thermodynamic limit,
and (ii) we minimize the vacuum energy of eqn. (\ref{w2}) with respect
to $A_{n},B_{n}$, and $\lambda$ on large finite lattices with $N\leqslant10^{4}$
sites and periodic boundary conditions. The results from both approaches
agree.

In the present work we set $S=1/2$ and study the ground state energy,
the quantum phases, and the spin correlation functions arising from
$A_{n},B_{n}$, and $\lambda$.

\subsection{Quantum Monte-Carlo\label{QMC-theory}}

We employ the stochastic series expansion (SSE) \cite{Sandvik1992,Sandvik1999a,Syljuaasen2002},
which is based on importance sampling of the high temperature series
expansion of the partition function \begin{equation}
Z=\sum_{\alpha}\sum_{n}\sum_{S_{n}}\frac{(-\beta)^{n}}{n!}\left\langle \alpha\right|\prod_{k=1}^{n}H_{a_{k},b_{k}}\left|\alpha\right\rangle \,\label{eq:partition}\end{equation}
where
$H_{1,b}=1/2-S_{i(b)}^{z}S_{j(b)}^{z}$ and $H_{2,b}=(S_{i(b)}^{+}S_{j(b)}^{-}+S_{i(b)}^{-}S_{j(b)}^{+})/2$
are spin diagonal and off-diagonal bond operators between sites $i,j$. $|\alpha\rangle=\left|S_{1}^{z},\ldots,S_{N}^{z}\right\rangle $
refers to the $S^{z}$ basis and $S_{n}=[a_{1},b_{1}][a_{2},b_{2}]\ldots[a_{n},b_{n}]$
is an index for the operator string $\prod_{k=1}^{n}H_{a_{k},b_{k}}$.
This string is Metropolis sampled using diagonal updates which change
the number of diagonal operators $H_{1,b_{k}}$ in the operator string
and directed loop updates which perform changes of the type $H_{1,b_{k}}\leftrightarrow H_{2,b_{k}}$.
For \emph{unfrustrated} spin-systems the latter update comprises an
even number of off-diagonal operators $H_{2,b_{k}}$, ensuring positiveness
of the transition probabilities.

Evaluation of the transverse dynamic structure factor with QMC is
performed in real space $i,j$ and at imaginary time $\tau$ following
Ref. \cite{Sandvik1992}
\begin{eqnarray}
S_{i,j}\left(\tau\right)=\Bigg\langle \sum_{p,m=0}^{n}\frac{\tau^{m}(\beta-\tau)^{n-m}n!}{\beta^{n}(n+1)(n-m)!m!}\times \phantom{aaa}\nonumber \\
S_{i}^{+}(p)S_{j}^{-}(p+m)\Bigg\rangle _{W}~,
\end{eqnarray}
 where $\langle\ldots\rangle_{W}$ refers to the Metropolis weight
of an operator string of length $n$ generated by the stochastic series
expansion of the partition function \cite{Sandvik1999a,Syljuaasen2002},
and $p,m$ are positions in this string. Analytic continuation to
real frequencies follows from the inversion of $S_{\perp}({\bf q},\tau)=\frac{1}{\pi}\int_{0}^{\infty}d\omega S_{\perp}({\bf q},\omega)K(\omega,\tau)$,
with a kernel $K(\omega,\tau)=e^{-\tau\omega}+e^{-(\beta-\tau)\omega}$
and $\beta=1/T$, and $S_{\perp}({\bf q},\tau)=\sum_{a}e^{i{\bf q}\cdot{\bf r}_{a}}S_{a,0}(\tau)/N$.

The preceding inversion is performed using the maximum entropy method
(MaxEnt), minimizing the functional $Q=\chi^{2}/2-\alpha\sigma$ \cite{bryan1,Jarrell1996}.
Here $\chi$ refers to the covariance of the QMC data to the MaxEnt
trial-spectrum $S_{\alpha\perp}({\bf q},\omega)$. Overfitting is
prevented by the entropy $\sigma=\sum_{\omega}S_{\alpha\perp}({\bf q},\omega)\ln[S_{\alpha\perp}({\bf q},\omega)/m(\omega)]$.
We have used a flat default model $m(\omega)$, matching the zeroth
moment of the trial spectrum. The optimal spectrum follows from the
weighted average of $S_{\alpha\perp}({\bf q},\omega)$ with the probability
distribution $P[\alpha|S({\bf q},\tau)]$ adopted from Ref. \cite{bryan1}.

\subsection{Exact diagonalization and density matrix renormalization group\label{DMRG-details}}

All DMRG and ED calculations employ the open source packages ALPS \cite{alps} and SPINPACK
\cite{spinpack}. We refer to their documentation.
In DMRG specifications,  $m$ refers to the number of states kept during sweeps.

\end{document}